# Train-borne Localization Exploiting Track-Geometry Constraints – A Practical Evaluation

Hanno Winter[1], Volker Willert[1], and Jürgen Adamy[1]

All authors: Control Methods and Robotics Laboratory, TU Darmstadt, Germany

**Abstract**

Today's railway signalling system heavily relies on trackside infrastructure such as axle counters and track balises. This system has proven itself to be reliable, however, it is not very efficient and, moreover, very costly. Thus, it is not suited to overcome the future challenges in railway transportation. For this reason, signalling systems based on train-borne sensors have gained interest recently. In this context, train-borne localization is one of the main research challenges. So far there is no sensor set-up which meets the demanding requirements for a localization system, both in the sense of accuracy as well as safety. To help overcome these issues in the near future we present our latest research results here. Earlier we published a localization algorithm which is characterized by an increased accuracy in cross-track direction compared to a standard Kalman filter (KF) approach, as has been shown in simulations [1]. To verify these results practically, we recorded data from a Global Navigation Satellite System (GNSS) and an inertial measurement unit (IMU) on a test drive. The localization accuracy is evaluated with the help of OpenStreetMap (OSM) data and site plans. Furthermore, we evaluate the quality of the estimated geometric track-map, which is additionally provided in the process of the localization algorithm [2]. We conclude with some remarks on the research challenges towards train-borne localization and suggest further steps to overcome them.

## 1 Introduction

To manage the constantly increasing traffic volume in the railway system it is necessary to utilize the existing infrastructure much more efficiently. Train-borne localization can be a key to achieve this. It allows for an optimal utilization of the existing infrastructure and at the same time many track-side elements, e.g. track-side signals, could be omitted. By that train-borne localization systems help to make the whole railway system not only more efficient in an operational but also in an economical sense.

However, there is no train-borne localization system available on the market yet [3]. This is due to the demanding safety requirements such a system has to fulfill according to EN 50126 [4]. The main challenge is to ensure a reliable and always available track-selective localization result, i.e. ±1.5m in cross-track direction [5]. Up to now, none of the investigated sensor set-ups could achieve the postulated safety requirements at such a positioning accuracy level. This is why train-borne localization has gained interest in research and development recently [3,6].

To help overcome the issues related to train-borne localization, on the one hand, we focus on developing new methods which help to improve the availability and accuracy of train-borne localization systems. On the other hand, we generate digital track-maps specifically tailored for train-borne localization. This is due to the fact that maps pose a single point-of-failure in the overall localization process. Therefore, we deem it is helpful to integrate a mapping method directly into the localization process which makes it possible to continuously provide accurate maps and to detect possible mapping errors.

[1] {hanno.winter,vwillert,adamy}@rmr.tu-darmstadt.de

Consequently, we presented a new localization approach in [1]. It is characterized by an increased positioning accuracy especially in bad GNSS situations and in cross-track direction. Additionally, we appended this approach by a mapping functionality which creates compact geometric track-maps being advantageous for train-borne localization applications [2].

Until now both approaches have only been evaluated with the help of simulations. In contrast to that, here we want to present an evaluation with the help of real measurement data.

## 2  Basics

The work presented in the remainder of this paper is based on two previous publications of ours. In [1] we presented a new train-borne localization filter and in [2] we extended it with a mapping function. In both, the working principal is demonstrated in simulations. Here we want to focus on the evaluation of the approaches with real measurement data. Prior to that, the most important ideas of the localization and mapping filter as well as the utilized sensor set-up will be presented briefly.

### 2.1  Localization and Mapping Filter

Our localization and mapping filter is characterized by an improved positioning accuracy, especially in bad GNSS situations and in cross-track direction. This is achieved by directly incorporating track-geometry constraints in the sensor-fusion process of the filter. The unique feature is that the track-geometry constraints are identified in real time [1]. Thus, no initial digital track-map is needed. Furthermore, the identified track-geometries are used to generate digital track-maps. Compared to the mostly used track maps, our maps are very compact, even though they contain additional track-geometry information [2]. The function principal of the filter is illustrated in Figure 1.

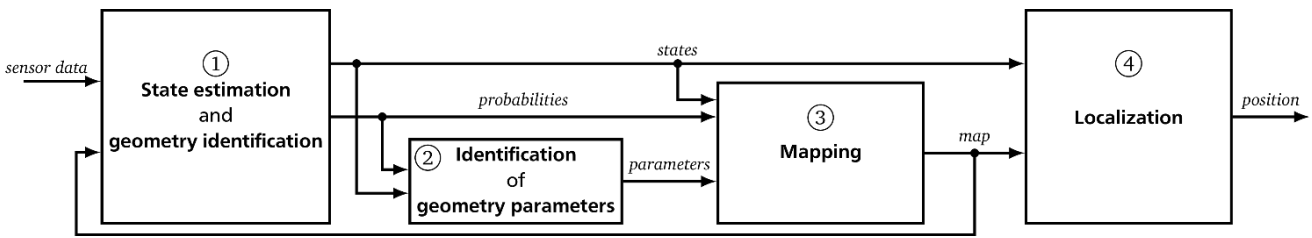

Figure 1: Working principal of the localization and mapping filter.

The first step (1) is to identify the track geometry the train is currently moving on, e.g. a straight or a circular arc track-segment. At the same time the kinematic states of the train, e.g. its current speed or heading angle are estimated. Therefore, an interacting multiple model (IMM) filter with three models is used. Each model is adapted to the train's motion on a particular track-geometry, i.e. motion on a straight line, motion on a circular arc or neither of them. In our simulations we assumed GNSS and IMU data to be available as input data, since at least these sensors can be considered to be available in any kind of train-borne localization system of the future [3,6]. This is also the kind of data which is available for the evaluation in the remainder of this paper. However, if available, it would make sense to integrate more sensor data in the filter, e.g. additional odometer or a speed data. In the next step (2) the state estimates and the geometry information are used to calculate the track's current geometry parameters, e.g. the starting point and elevation of a straight line. Then (3) the current parameters are combined with all previously identified track-geometries to generate a compact geometric track-map. Last (4), the map and the state estimates are combined to a position estimate. Once a track-map has been generated it can serve as additional input for future journeys on this track. By that the generated map and the achievable positioning accuracy will improve over time.

## 2.2 Measurement Set-Up

The underlying data for this paper originate from a test drive between Annaberg-Buchholz and Schwarzenberg in the Erzgebirge in Germany. They were recorded at the 24-th of October in 2018 under cloudy weather conditions. The track is visualized in Figure 2. It is a non-electrified secondary line in a harsh railway environment, i. e. tight curves, steep slopes, forested embankment and strongly changing weather conditions. The track is not used regularly but still fully maintained. As a result, this track is often used for testing new technologies. The following evaluations were carried out on the section red marked in Figure 2.

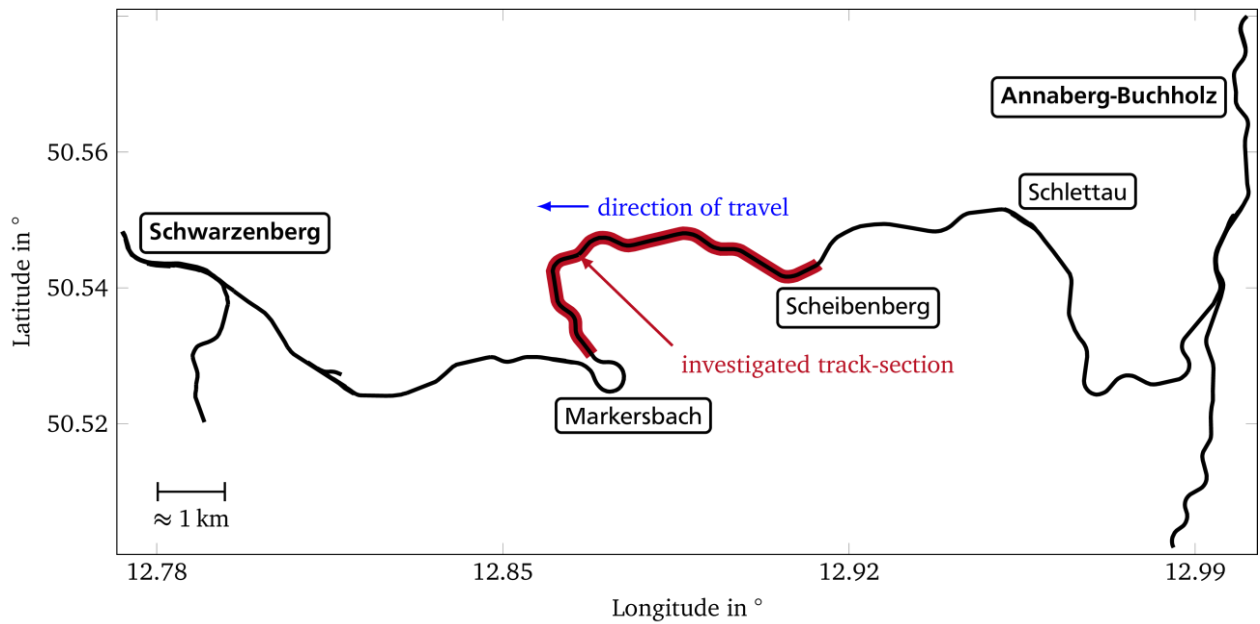

Figure 2: Overview of the track between Annaberg-Buchholz and Schwarzenberg.

It is the 5.7 km long section from the railroad crossing in Scheibenberg to the end of the Markersbacher viaduct. All data for this paper was recorded with an iMAR iNAT-M200/STN sensor. It is a powerful MEMS based INS/GNSS navigation system. Here only the raw GNSS- and IMU-data[2] provided at a rate of 1Hz and 500Hz respectively were used to calculate a separate navigation solution with our algorithm. The GNSS antenna was installed at a roof platform at the front of the train as shown in Figure 3. The IMU was installed in the passenger cabin near the front doors, a bit in front of the bogie and slightly shifted to the right in the direction of travel (c.f. Figure 4).

The combination of GNSS and IMU is rather common in navigation applications as they complement each other very well. Most of the time the GNSS provides an absolute positioning solution which can be continued during GNSS outages with the help of the IMU data. Since the IMU data are drift afflicted the accuracy of the positioning solution decreases with time. Therefore, only GNSS outages of a few seconds (depending on the quality of the utilized IMU) can be tolerated in a train-borne localization system. Despite that, the combination of GNSS and IMU can be said to be one certain component of train-borne localization systems of the future which has to be supplemented by additional sensors [3,6]. Although reasonable, no additional sensors are considered since there were none available. However, the following evaluation will demonstrate how our algorithm helps to increase the positioning accuracy of a solely IMU/GNSS based localization system especially in bad GNSS situations and in cross-track direction.

---

[2] More precisely, the position, velocity and time (PVT) solution of the GNSS receiver, as well as the 6-DOF IMU (accelerations and turn rates) measurements, are used.

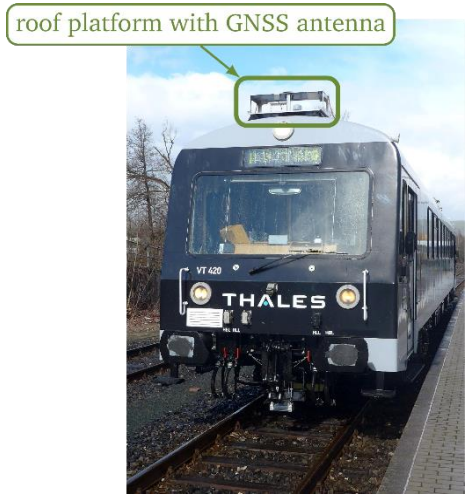

Figure 3: Front view of the test vehicle. The GNSS-Antenna was mounted on the marked roof platform.

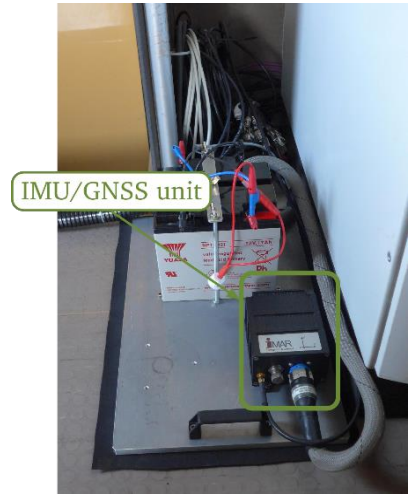

Figure 4: Measurement platform in the passenger cabin. The IMU/GNSS unit can be seen in the marked region.

## 3  Analysis

In this section we evaluate our localization and mapping algorithm [1,2]. The evaluation is separated in three steps: First, some raw data are presented to get a feeling of the situation at hand and what kind of quality can be expected of the localization and mapping filter. Second, the localization results are evaluated and third, the mapping results are evaluated.

### 3.1  Sensor Data Preview

To get a feeling of the situation at hand some measurements which are thought to have the biggest influence on the localization and mapping results are shown in Figure 5. The presented measurements are the GNSS positioning error in terms of the standard deviation $\sigma$, the GNSS speed and the yaw rate of the vehicle. The first two measurements are provided by the GNSS receiver whereas the yaw rate is provided by the IMU.

In the top diagram of Figure 5 the maximum $3\sigma$ GNSS positioning error related to a confidence level of 99.73% is presented. This error corresponds to a confidence level of 99.73% (assuming a normally distributed error) for the real position to be within the interval of $\pm 3\sigma$ around the position estimate. The minimum $3\sigma$ error is 3.4m and the mean $3\sigma$ error is 25.3m. A valid GNSS solution was available around 95% of the time. The longest GNSS outage of 13s occurred at 268 – 281s. The degradation and losses of the GNSS signal mostly occur in places with dense forest around the track. Although this performance may seem poor compared to the absolute requirements in the sense of accuracy and availability for train-borne localization systems [7], it gives a good impression of the performance that can be expected of GNSS without additional correction data. It is worth to notice that the following evaluation of the localization performance is limited to the quality of these GNSS measurements as there are no better reference measurements available for this test drive. This is a general problem when evaluating train-borne localization systems. For this reason, we and others recommend to carry out more test drives and, by that, help to create extensive reference datasets for the evaluation of train-borne localization systems [6,8].

The diagram in the middle of Figure 5 shows the speed measured by the GNSS receiver. It varies between 22km/h and 56km/h (neglecting the outliers around the main GNSS outage at 268s). This speed profile is typical, as it ranges from medium speeds to the maximum speed allowed on this track.

The bottom diagram of Figure 5 shows the yaw rate of the vehicle. The data is available at all time since it is provided by the IMU. On straight track-elements the yaw rate varies around zero whereas it is non-zero on curved track-elements. This behavior can be observed very well in the course of the yaw rate as there are clear variations from zero. This can make it plausible how the investigated filter can identify the current track-geometries online, although the yaw rate is not the only feature utilized for this purpose in the filter.

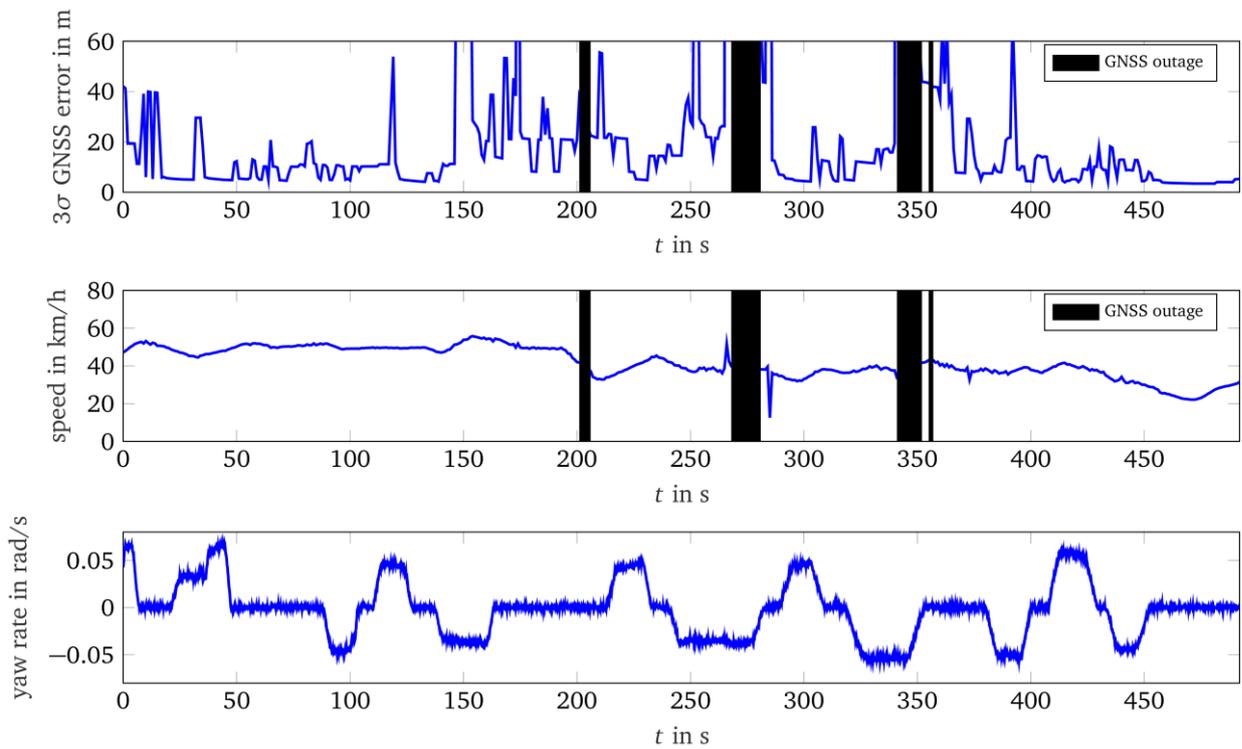

Figure 5: Raw sensor data of the GNSS receiver (top and center plot) and the IMU (bottom plot). Top: 3σ GNSS positioning error (corresponding to a confidence level of 99.73%) provided by the sensor itself. Center: Speed measurement provided by the GNSS receiver. Bottom: Yaw rate provided by the IMU.

## 3.2 Localization

At first the localization accuracy is evaluated in the sense of the 3σ positioning error. In total three different positioning approaches are compared. These are the GNSS positioning solution, the sensor fusion result obtained with a standard Kalman filter (KF) approach and the sensor fusion result of our algorithm presented in [1]. The 3σ positioning error is calculated based on the covariance matrices provided by either the GNSS receiver or the sensor fusion filters.

In Figure 6 the cumulative distribution function (CDF) of the 3σ positioning error is shown. It shows the availability of the positioning result over its accuracy. This makes it possible to easily compare the performance of the different localization approaches in the sense of accuracy and availability. Three different CDF plots are shown. From the left to the right these are the maximum 3σ positioning error

(left), the error in along-track direction (middle) and the error in cross-track direction (right). Some interesting data points of these diagrams are summarized in Table 1 and Table 2.

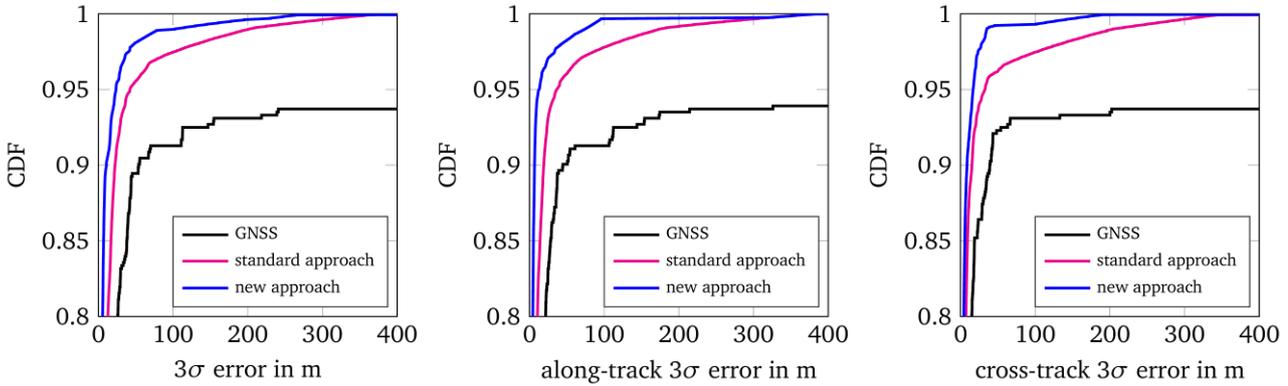

Figure 6: Cumulative distribution function (CDF) of the $3\sigma$ positioning error (corresponding to a confidence level of 99.73%). The error is calculated from the covariance matrices provided by either the GNSS receiver or the sensor fusion algorithms.

By the plots the preliminary simulation results can be confirmed also for real measurement data. The new algorithm clearly increases the positioning accuracy compared to the standard KF sensor fusion approach. This is especially true for the error in cross-track direction (c.f. Figure 6 and Table 1). A comparison of, e.g. the 0.99 CDF values (c.f. Table 1) shows that the new approach improves the positioning accuracy in along-track direction by approximately 55% and in cross-track direction by approximately 80% compared to the standard KF approach. A solely GNSS based positioning solution is not available for a CDF value of 0.99 due to the GNSS outages. In absolute numbers a GNSS positioning solution is only available for approximately 95% of the time, but this also includes very inaccurate positioning solutions with a $3\sigma$ error of more than 400 meters (c.f. Figure 6). In contrast to that the sensor fusion approaches are available all the time, which can be seen in the CDF plots due to the fact that both curves reach a CDF value of 1. Admittedly, this happens only for poor $3\sigma$ positioning errors (c.f. Table 1).

An exemplary comparison of the positioning solutions at fixed $3\sigma$ errors of respectively ±5m in along-track and ±1.5m in cross-track direction is shown Table 2. These values are either motivated by the allowed error for the localization with balises in ETCS, ±5m, or the theoretical value of ±1.5m to enable a track-selective localization. Again, the new algorithm shows an increased performance for both error values. The new algorithm at least allows a track-selective localization in 25% of the time (at a confidence level of 99.73%) whereas this is impossible with the other two approaches.

A qualitative evaluation of the localization results is shown in Figure 7. It shows the localization results of the standard KF approach and the new approach together with their corresponding $3\sigma$ error-ellipses in a phase with reduced GNSS performance. Additionally, the GNSS positioning results are shown but without their error-ellipses for the sake of clarity. All results are mapped on a satellite image.

The ellipses of the new approach are significantly smaller than the ellipses of the standard approach. This is a direct result of the track-geometry information which is incorporated in the new approach. It leads to a more constrained positioning uncertainty in cross-track direction. As this information is not available in the standard approach the GNSS uncertainty effects the localization result much more. As a reference, the track data from OpenStreetMap (OSM) [9] is plotted in Figure 7 additionally. From that it can be seen that neither the GNSS measurements nor the localization results of both sensor fusion approaches lie near the OSM track. Besides that, the OSM track does not even match with the track visualized on the satellite image. Therefore, it can be suspected that the OSM track is not very accurate in the viewed section. This may be important to have in mind throughout the following evaluation of the mapping results.

Table 1: 3σ positioning error (corresponding to a confidence level of 99.73%) at specific values of the cumulative distribution function (CDF).

|     | CDF  | 3σ error in m |          |       |
|-----|------|---------------|----------|-------|
|     |      | GNSS          | standard | new   |
| AT[3] | 0.90 | 44.6          | 19.2     | 6.9   |
| CT[4] | 0.90 | 39.1          | 15.2     | 8.7   |
| AT  | 0.99 | inf           | 182.9    | 76.5  |
| CT  | 0.99 | inf           | 204.5    | 35.3  |
| AT  | 1.00 | inf           | 382.9    | 377.7 |
| CT  | 1.00 | inf           | 340.2    | 188.3 |

Table 2: Cumulative distribution function (CDF) at specific values of the 3σ positioning error (corresponding to a confidence level of 99.73%).

|    | 3σ error in m | CDF   |          |       |
|----|---------------|-------|----------|-------|
|    |               | GNSS  | standard | new   |
| AT | ±5m           | 0.314 | 0.523    | 0.833 |
| CT | ±1.5m         | 0.000 | 0.058    | 0.259 |

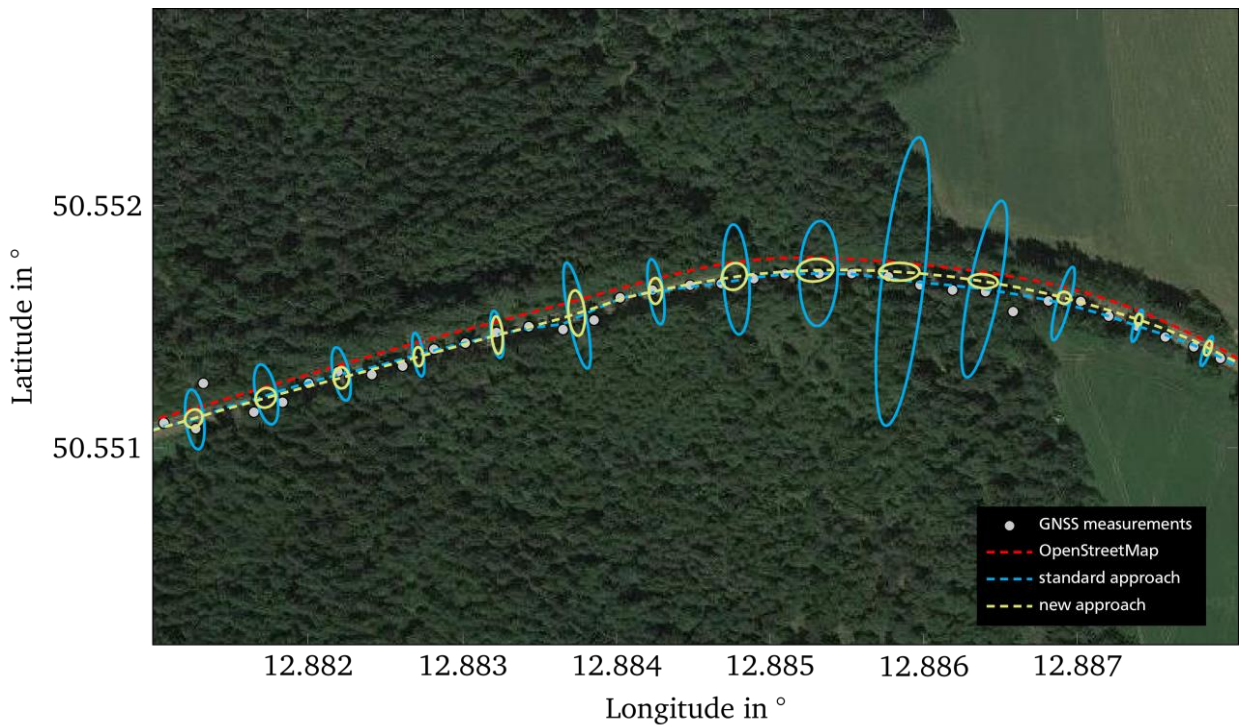

Figure 7: Visualization of the different positioning accuracies on a satellite image (Image © 2019 Google, Maps © 2019 Geo-Basis-DE/BKG (© 2009), Google) during a phase of reduced GNSS performance. By comparison of the sizes of the 3σ error ellipses (corresponding to a confidence level of 99.73%) the increased accuracy in cross-track direction of the new approach can be seen clearly.

---

[3] AT: along-track

[4] CT: cross-track

## 3.3 Mapping

Next, the quality of the map generated with the new algorithm (c.f. Section 2.1 and [2]) will be examined. Therefore, the results of the track-geometry identification process and the finally generated map are visualized on a satellite image in Figure 8.

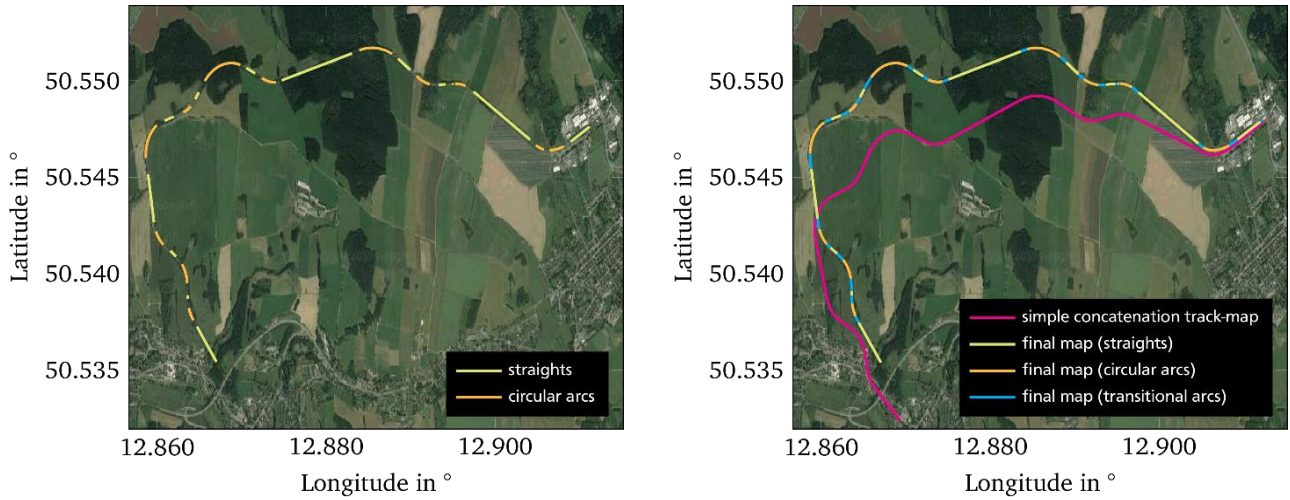

Figure 8: Visualization of the track-geometry identification process (left) and the finally generated map (right) on a satellite image (Image © 2019 Google, Maps © 2019 GeoBasis-DE/BKG (© 2009), Google).

Left: Visualization of the results of the second step in the localization and mapping filter (geometry-parameter identification). At this step only the parameters of straights and circular arcs are fully identified. Finding suitable parameters for the connecting transitional arcs is the task of the next step.

Right: Visualization of the mapping result. The final map consists of a continuous sequence of geometric track-elements, describing the real course of the track very well. The poor initial map results from the simple concatenation of the identified track-geometries which have been identified by the localization filter.

The online identified track-geometry parameters are those of straights and circular-arcs. The resulting track elements are shown in Figure 8 (left). It can be seen that all main straights and circular-arcs were identified quite accurate. This confirms the simulation results once again, i.e. the track-geometry parameter identification also works on real measurement data. In most cases the sequence "straight – gap – circular-arc – gap – straight" is present. Two times a sequence of "circular-arc – gap – circular-arc" occurs. This is best visible at the beginning of the track (take the direction of travel from east to west into account) in the first right turn. This is not a misdetection. Taking a look at the yaw-rate diagram in Figure 5 between $20 - 50s$ makes it plausible that at this point indeed two circular-arcs with different radii are following after each other. This is plausible due to the yaw-rate being at one non-zero level between $20 - 35s$ and then changing to a higher level between $35 - 50s$ without first returning to zero. The second time a "circular-arc – gap – circular-arc" sequence occurs at $100 - 105s$. Here the yaw-rate curve is not that clear. Only a small level-changing behavior can be seen. Thus, without construction plans it is difficult to tell if a misdetection has occurred or not. But for the following processing of the identified track-elements it is better to have several small track elements representing the true course of the track very well than having few track-elements with a rather poor fitting quality. In this sense it can be noted once again that the track-geometry parameter identification works well.

In the next step (c.f. Section 2.1) the identified track-elements are combined to a continuous total track. Therefore, the gaps between the identified track-elements are assumed to be transitional-arcs realized as clothoids. This can be assumed because of the construction principals of railway tracks [10]. Theoretically all parameters determining the connecting clothoids are already known from the localization and mapping filter. These parameters are: Starting point, starting direction, starting respectively ending radius of the clothoid and its length. However, the simple concatenation of all track-elements including the clothoids gives a rather poor mapping result which can be seen in Figure 8 (right). The problem is, that even small deviations in one of the first track-element's parameters have a huge impact on the position of all following track-elements due to the continuous concatenation of all tracks. Therefore, an optimization strategy is applied to find a good fit of all track-element parameters into the known positioning results [2]. The result is also shown in Figure 8 (right). It can be seen that the track matches with the course on the satellite image in the background.

The final map parameters representing the whole track can be stored in a very compact form as exemplary shown in Table 3. This compact representation is worth the initial effort as it has some major advantages compared to the often used data-point based map representations. One advantage is its compactness itself, resulting in huge memory savings which in turn have many benefits, e. g. in wireless applications or situations where the map has to be searched through. Moreover, all geometric parameters of the track are directly accessible and can be used easily for further calculations, e. g. the calculation of the perpendicular distance of a point to the track.

Table 3: Excerpt of the identified compact geometric track-map.

| Track element number | 1 | 2 | 3 | 4 | 5 | 6 | 7 | 8 | 9 | 10 | 11 | ... | 53 |
|---|---|---|---|---|---|---|---|---|---|---|---|---|---|
| Shape[5] | st | ta | ca | ta | st | ta | ca | ta | ca | ta | st | ... | st |
| Length $L$ in m | 0 | 11 | 27 | 25 | 218 | 76 | 136 | 37 | 87 | 48 | 581 | ... | 296 |
| Radius $r$ in m | $\infty$ | 213 | 213 | 213 | $\infty$ | 376 | 376 | 197 | 197 | 197 | $\infty$ | ... | $\infty$ |
| Starting point | \multicolumn{6}{c}{$p_0 = [50.548° \; 12.913°]^T$} | | | | | \multicolumn{3}{c}{$\psi_0 = 231.2°$} | | |

From the previous qualitative evaluation, it can be stated that our mapping algorithm also works in practice as it did in the simulations. Despite that, next a quantitative evaluation of the mapping result should be carried out with the help of a reference track extracted from OSM data. In Figure 9 the CDF of the absolute mapping error $|\varepsilon|$ in perpendicular direction from the calculated map to the OSM map is shown. Additionally, some significant values of $|\varepsilon|$ are listed in Table 4.

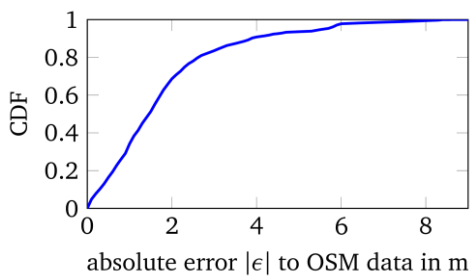

Figure 9: Cumulative distribution function (CDF) of the absolute error $|\varepsilon|$ in meters between the calculated map and the map from OpenStreetMap (OSM).

Table 4: Absolute error $|\varepsilon|$ in meters between the calculated map and the map from OpenStreetMap (OSM).

|  | mean | \multicolumn{3}{c}{CDF} | | |
|---|---|---|---|---|
|  |  | 0.95 | 0.99 | 1.00 |
| $|\varepsilon|$ in m | 1.80 | 5.7 | 7.5 | 8.7 |

The mean error of the calculated map to the OSM map is 1.8m and the maximum error is 8.7 m (c.f. Figure 9 and Table 4). These values confirm that the calculated map is quite accurate. However, as mentioned above, also the OSM map can contain inaccuracies which make it difficult to generalize from the presented errors.

---

[5] st: straight, ta: transitional arc, ca: circular arc

### 3.4 Concluding Remarks for Train-Borne Localization Systems

Based on the above analyses we want to make some remarks concerning further steps when developing train-borne localization systems of the future.

We examined the localization quality in the sense of the 3σ positioning error. This corresponds to a confidence level of 99.73% for the real position to be within a range of ±3σ. However, this confidence level is by far not acceptable for safety critical railway applications in the common sense of EN 50126 [4]. Therefore, presumably the 6σ positioning error would have had to be considered. If this confidence level had been considered here, none of the here presented localization techniques would have provided a track-selective localization result ever. That raises the question if it is even possible to achieve a track-selective localization result with any combination of the available sensors for train-borne localization systems under the current circumstances of EN 50126 [4].

We think that it is very likely impossible to achieve a valid safety prove by mainly focusing on the sensor configuration for train-borne localization systems. Instead also new methods to assess the safety of train-borne localization systems and the dynamic overall systems associated with them should be investigated further. This becomes clearer with the following explanations: In the above statements it was implicitly assumed that a track-selective localization has to be possible based on a single localization result. This must not necessarily be the case as the determination of the correct track can also be treated as a dynamic process. Additionally, as the uncertainty of the positioning solution could be estimated at any time, it would make sense to dynamically account for it in the current movement authorities of all trains. This principal would enable a safe operation independent of the absolute size of the current uncertainties. By that it becomes plausible that it makes sense to develop customized methods and measures for a safety prove of train-borne localization systems.

Additionally, in order to develop suitable safety methods and safety measures we think it is essential to carry out much more real tests like the one presented here. Only by that all the short and long-term effects influencing the performance of train-borne localization systems can be identified and treated accordingly. Therefore, we and others already recommended to create extensive reference datasets for the evaluation of train-borne localization systems [6, 8].

## 4 Summary

In this paper we investigated the performance of a fully train-borne localization and mapping filter we presented earlier. Until now the performance has only been investigated using simulations. In contrast to that, in this work the performance has been evaluated with the help of real GNSS and IMU measurement data collected on a test drive. By that, our preliminary simulation results could be confirmed. The new approach is capable of increasing the localization accuracy which is especially true for the accuracy in cross-track direction and during bad GNSS situations. Moreover, an accurate compact geometric track-map could be generated confirming the preliminary simulation results once again. Beside the evaluation of our localization and mapping filter we gave some concluding remarks concerning the further development of train-borne localization systems of the future. According to that we recognized a great need for developing new safety methods specifically designed for train-borne localization systems. Additionally, we suggest to carry out much more real tests and, by that, to build extensive reference datasets for the evaluation of train-borne localization systems.

## 5 Acknowledgements

We kindly thank DB Netz AG for supporting this research project. Furthermore, we like to thank Thales affording us to collect the raw data with their test vehicle LUCY, and the group of Geodetic Measurement Systems and Sensors at TU Darmstadt for providing the IMU/GNSS sensor platform.